\documentclass[reprint, english, amsmath,amssymb,aps,nofootinbib, prd]{revtex4-1}

\usepackage[T1]{fontenc}
\usepackage[latin9]{inputenc}
\usepackage{color}
\usepackage{graphicx}
\usepackage{amssymb,amsmath}
\usepackage{babel}

\def\be{\begin{equation}}
\def\ee{\end{equation}}

\makeatletter

\@ifundefined{definecolor}
{\usepackage{color}}{}

\newcommand{\beq}{\begin{equation}}
\newcommand{\eeq}{\end{equation}}  
\newcommand{\ba}{\begin{eqnarray}}
\newcommand{\ea}{\end{eqnarray}}
\newcommand{\bef}{\begin{figure}}
	\newcommand{\eef}{\end{figure}}

\newcommand{\eff}{\text{eff}}

\newcommand{\nn}{\nonumber}

\newcommand{\dn}{\Delta N_\text{eff} }

\begin{document}
	
	\title{Observable windows for the QCD  axion through $N_\eff$}
	
	\author{Ricardo Z. Ferreira$^{1}$}
	\email{rferreira@icc.ub.edu}
	\author{Alessio Notari$^{1}$}
	\email{notari@fqa.ub.edu}
	
	\affiliation{$^{1}$ Departament de F\'isica Qu\`antica i Astrofis\'ica \& Institut de Ci\`encies del Cosmos (ICCUB), Universitat de Barcelona, Mart\'i i Franqu\`es 1, 08028 Barcelona, Spain}

	\begin{abstract}
		
We show that when the QCD axion is directly coupled to quarks with $c_i/f \, \partial_\mu a \, \bar{q}_i \gamma^\mu \gamma^5 q_i$, such as in DFSZ models, the dominant production mechanism in the early universe at temperatures $1 \, {\rm GeV}\lesssim T \lesssim 100 \,{\rm GeV}$ is obtained via $q_i \bar{q}_i \leftrightarrow g a$ and $q_i  g \leftrightarrow q_i a$, where $g$ are gluons.  The production of axions through such processes is maximal around $T\approx m_i$, where $m_i$ are the different heavy quark masses. This leads to a relic axion background that decouples at such temperatures, leaving a contribution to the effective number of relativistic degrees of freedom, which can be larger than the case of decoupling happens the electroweak phase transition, $\Delta N_{\rm eff}\lesssim 0.027$. 
Our prediction for the $t$-quark is $0.027\leq \dn \leq 0.036$ for $10^6 {\rm GeV} \lesssim f/c_t\lesssim 4\times 10^8$ GeV and for the $b$-quark is $0.027\leq \dn \leq 0.047$ for $10^7 {\rm GeV}  \lesssim f/c_b\lesssim 3\times  10^8  {\rm GeV}$. For the $c$-quark the window can only be roughly estimated as $0.027<\dn\lesssim {\cal O}(0.1)$, for $f/c_c\lesssim (2-3) \times 10^8  {\rm GeV}$, since axions can still be partially produced in a regime of strong coupling, when $\alpha_s\gtrsim 1$.  
%If the axion is not fully thermalized smaller values of $\dn$ would be obtained. 
These contributions are comparable to the sensitivity of future CMB S4 experiments, thus opening an alternative window to detect the axion and to test the early universe at such temperatures. 

	\end{abstract}

	\maketitle

	\section{Introduction}
	
	The axion ($a$) is a field postulated for theoretical reasons~\cite{Peccei:1977ur,Peccei:1977hh}, that possesses a very rich phenomenology~\cite{Patrignani:2016xqp}. 
	It has a coupling to gluons of the form $\alpha_s /(8 \pi f)\,  a G \tilde{G}$, that can provide a compelling solution to the strong CP problem~\cite{Cheng:1987gp}, {\it i.e.} the fact that the total coefficient of the CP violating operator $\theta/(8\pi) G\tilde{G}$ has a coefficient consistent with zero in vacuum with high precision, $|\theta|\lesssim 10^{-10}$~\cite{Baker:2006ts,Afach:2015sja}. Experimentally the possible range for $f$ is bounded by $f\gtrsim 10^6-10^9$ GeV, via direct detection experiments~\cite{Patrignani:2016xqp} or stellar cooling bounds~\cite{Hansen:2015lqa, Feng:1997tn}~\footnote{In the case of a coupling to the top quark one gets even $f\gtrsim {\cal O}(10^9)$GeV~\cite{Feng:1997tn}.}, although different bounds refer to coupling to different particles, namely photons, nucleons, quarks or leptons. From Supernova SN1987A~\cite{Patrignani:2016xqp,Raffelt:2006cw,Fischer:2016cyd,Giannotti:2017hny} one gets $f\gtrsim {\cal O}(10^8)$ GeV, but such bound should be considered less robust due to complicated SN physics. Moreover, the SN bound relies specifically on the axion-nucleons couplings and so is somehow model dependent. In fact the low energy effective Lagrangian for the QCD axion can arise from several UV completions, with different properties, leading thus to different effective couplings to the various Standard Model (SM) fields, including quarks and nucleons.
The coherent oscillations of the axion can lead also to a good candidate for Dark Matter, for $f\gtrsim 10^{11}$ GeV~\cite{Patrignani:2016xqp, Preskill:1982cy, Abbott:1982af, Dine:1982ah}.

Axions also naturally constitute a population of detectable hot relics from the early universe~\cite{Turner:1986tb,Berezhiani:1992rk,Chang:1993gm,Masso:2002np,Graf:2010tv,Salvio:2013iaa,Brust:2013xpv} if the interactions that can produce them are fast enough at some stage of the radiation dominated universe. A rather generic expectation is that axions can be produced in thermal abundance either via interactions with gluons~\cite{Masso:2002np,Graf:2010tv}, or via direct interactions with the top quark~\cite{Turner:1986tb,Salvio:2013iaa}, and they subsequently decouple and freeze out at a high temperature $T_D$, before the Electroweak phase transition (EWPT).

Their relic abundance in this case would be proportional to the inverse of the effective number of degrees of freedom at the decoupling temperature $g_{*, D}$ and this would affect the energy density of relativistic particles at recombination, traditionally parameterized by an effective number of neutrinos $N_{\rm eff}$, which can be tested by the CMB.  
Assuming only the existence of the SM plus the axion above the EWPT, the predicted change in $N_{\eff}$ due to axions would be given by $\Delta N_\eff = 4/7 \, (43/(4g_{*, D}))^{4/3} \approx 0.027$ \cite{Kolb:1990vq}. Remarkably this is comparable to the forecasted sensitivity of future CMB-S4 experiments~\cite{Baumann:2016wac, Abazajian:2016yjj}. If there are additional degrees of freedom beyond the SM at such high temperature $T_D$, this number would be smaller, so that this value is actually an upper bound. A possible thermalization at much later stages, below the QCD phase transition (QCDPT), via scattering with pions and nucleons, has been studied in~\cite{Berezhiani:1992rk, Chang:1993gm}, concluding that this would require $f\lesssim {\cal O}(10^6)$ GeV, which is already in conflict with the astrophysical bounds~\cite{Brust:2013xpv}.

In this {\it Letter} we point out that if the axion  is produced instead between the EWPT and the QCDPT,  it can erase the previous prediction due to the high temperature thermalization before the EWPT, leading instead to larger values of $\Delta N_\text{\eff}$, thus slightly easier to detect. Moreover, although there is room to have extra relativistic species at temperatures between the QCDPT and the EWPT, {\it e.g.} light relics or other weakly coupled particles, so far the Standard Model has passed precision tests with great success and so in this window of temperatures the values of $\Delta N_{\eff}$ can be seen more confidently as true predictions, as a function of $f$, instead of upper bounds. This happens due to direct couplings to various heavy quarks which are present in many axion models, as estimated already with dimensional analysis in~\cite{Turner:1986tb}. Similar couplings were also studied in~\cite{Cadamuro:2010cz} and~\cite{Baumann:2016wac}, although this was applied mostly to leptons, whose cross-sections are suppressed due to the absence of coupling to gluons. We explicitly compute here the relevant cross-sections, with $t$-$b$ and $c$ quarks,  at temperatures sufficiently above the QCDPT, so that we can use a perturbative approach, which should be roughly correct as long as $\alpha_s\lesssim 1$. This calculation is timely because of the planned CMB-S4 experiments, which should be able to go close to the necessary sensitivity.

	\section{Axial Couplings}
	
	The axion has a Lagrangian that includes couplings with gauge bosons and can also include direct derivative couplings with fermions. We are in particular interested in the direct interaction with the different quarks $q_i$ of the SM,
	\begin{eqnarray} \label{Lagrangian}
	{\cal L}_{int}= \frac{\partial_\mu a}{2f} \sum_i c_i \bar{q}_i \gamma^\mu \gamma^5 q_i \, , \label{directcoupling}
	\end{eqnarray}
	The coefficients $c_i$ are non-zero already at tree level, for example, in DFSZ models \cite{Zhitnitsky:1980tq,Dine:1981rt} and are typically of order unity~\footnote{We do not consider here the case of couplings that mix different flavors. This has been considered in~\cite{Baumann:2016wac}.}. Note also that in UV-completed axion models the $c_i$ are proportional to the Peccei-Quinn charges over the domain wall number $N_{DW}$~\cite{Kolb:1990vq}. 
	At high temperature, above the EWPT, this type of couplings are the most efficient ones that can thermalize the axion~\cite{Salvio:2013iaa}, through the scatterings $t+ h \rightarrow t +a,  \, \bar{t}+ h \rightarrow \bar{t} +a, \, t+ \bar{t} \rightarrow h +a,$ where $h$ is the Higgs particle.
	This is even larger than the scattering with gluons because it is not suppressed by $\alpha_s/(2\pi)$, contrary to the axion-gluon coupling, jointly with the fact that the top Yukawa coupling is $y_t \simeq 1$. By dimensional analysis the rate of the axion-top-Higgs process goes as $\Gamma_{t-h} \propto y_t^2 c_t^2 T^3/f^2$, and so can be larger than the Hubble rate   $H= \sqrt{g_*/90} \pi T^2/M_p$, at sufficiently high temperature, leading to an equilibrium distribution and subsequent decoupling when $\Gamma_{t-h}\simeq H$. Here, $M_p=2.4 \times 10^{18}$ GeV. Such a rate is larger than the one through gluons, which is given parametrically by  $\Gamma_g \propto (\alpha_s/(2\pi))^2 g_s^2  T^3/f^2$, where $g_s\equiv \sqrt{4 \pi \alpha_s}$.
	
	In this {\it Letter}, we will instead look at temperatures below the EWPT. At these energies the Higgs particles are Boltzmann suppressed, so a much more relevant scattering is obtained replacing the Higgs with a gluon, jointly with the fact that $g_s\gtrsim 1$. So we will consider the processes
	\begin{eqnarray} \label{Scatterings}
	&q_i+ g \rightarrow q_i+ a,  \quad \bar{q}_i+ g \rightarrow \bar{q}_i+ a&  \, ,\nn \\
	& q_i+ \bar{q}_i \rightarrow g+ a \, . 
	\end{eqnarray}
	A crucial feature of  these cross sections is that, similarly to the case of $\Gamma_{t-h}$, they are proportional to the Yukawa coupling squared, $y_i^2$ , which means to the squared mass of the quark, since the Higgs takes a vev $v$ below the EWPT. This can be understood performing a field redefinition, as in~\cite{Salvio:2013iaa}, which erases the coupling of eq.~(\ref{directcoupling}) and replaces it with a coupling of the type $ y_i c_i/f \, a h \bar{q}_i \gamma^5 q_i$. When the Higgs has a vev,  one can set $h=v$ and so the quark mass  $(y_i v)^2/2=m_i^2$ will actually appear in the cross sections. We work in the present letter without performing such a redefinition, but we verify by explicit calculation that all cross sections indeed go as $m_i^2$.
	
For $T\gg m_i$ this represents a suppression. However, when $T\simeq m_i$ that is no longer the case, and such interactions dominate, {\it e.g.}, over the ones with gluons.
	
	For temperatures above the QCDPT, the cross sections $\sigma_i$ can be evaluated straightforwardly and they are rescaled for the various quarks $q_i$ simply by the respective $m_i^2$. 
	In the end we are interested in computing standard thermally averaged rates (see {\it e.g.}~\cite{Kolb:1990vq,Masso:2002np}),
	\begin{eqnarray}
	\Gamma_i = \frac{1}{n_{eq}} \int \frac{d^3 p_a d^3 p_b}{(2\pi)^6 4 E_a E_b}  f_a f_b  (\sigma_i 4E_a E_b) \, ,
	\end{eqnarray}
	where we integrate over the 3-momenta of the incoming states, $f_{a,b}$ are their thermal distributions (Bose-Einstein or Fermi-Dirac) and $n_{eq}$ is the axion equilibrium number density.
	Parametrically~\cite{Turner:1986tb} one has $\Gamma_i \propto m_i^2 c_i^2 g_s^2 T/f^2$, at $T\gtrsim m_i$. If, instead, $T\lesssim m_i$ a Boltzmann suppression arises, due to $f_a$ and $f_b$, leading to 
	\begin{eqnarray}\Gamma_i \propto 
	e^{-\frac{m_i}{T}} \frac{m_i^2 c_i^2 g_s^2 T}{f^2} \, .
      	\end{eqnarray}
	
	In order to find the precise numerical pre-factor we explicitly compute the cross-sections. We simply use tree-level diagrams, adding also thermal masses in the kinematics; for a more refined treatment one could evaluate in thermal field theory the axion self-energy, as in \cite{Graf:2010tv, Salvio:2013iaa}, although we expect this will lead only to minor corrections.
	Comparing the scattering rates $\Gamma_i$ with $H$ allows us to know, at a given temperature, at which value of $f$ the axion is in thermal equilibrium. More precisely, we check whether $\Gamma_i/(3H) >1$. At a given $T$ the most important channel is the scattering with the heaviest quark which is not yet Boltzmann suppressed. Thus, the rates $\Gamma_i$ are maximal around $T\simeq m_i$.

	For the first two processes in eq.~(\ref{Scatterings}) the cross section can be quite simply expressed under the approximation of massless gluons ({\it i.e.} neglecting thermal masses) as 
	\begin{eqnarray} 
	4 \sigma^{(1)}_i &v& E_a E_b  = \frac{4 c_i^2 g_s^2 m_i^2}{\pi f^2}  \nn \\
	&& \times \left[ \tanh^{-1}\left(\frac p E \right) - \frac{p (2p +E)}{\left(p+E \right)^2} \right] \, ,
	\end{eqnarray}
	where $p$ and $E$ are the momentum and energy of the quark, respectively, in the center of mass frame.
	Instead, the cross section for pair annihilation takes a simple form even including the gluon thermal mass $m_{g,th}$,
\begin{eqnarray} 
4 \sigma^{(2)}_i &v& E_a E_b  = \frac{ 2c_i^2 g_s^2 m_i^2}{ \pi f^2 } \nn \\
	&& \times   \frac{4 E^2-m^2_{g,th}}{E \,p} \tanh^{-1} \left(\frac{p}{E}\right) \,.
\end{eqnarray}

In our final results we will include gluon thermal masses~\cite{Bellac:2011kqa},
\begin{eqnarray}
m^2_{g,th}=\left(\frac{N_c}{6} + \frac{N_f}{12} \right) g_s^2 T^2 \, ,
\end{eqnarray}
where $N_c$ and $N_f$ are respectively the number of colors and flavors.
We will also use in the quark dispersion relations an overall effective mass which includes both the standard mass $m_i\equiv y_i v/\sqrt{2}$ and the thermal mass \cite{Bellac:2011kqa}
\begin{eqnarray}
m^2_{i, {\rm eff}} = m^2_{i,th}+ m_{i}^2= \frac{N_f}{8} g_s^2 T^2 + m^2_{i} \, .
\end{eqnarray} 
Note however that, while the thermal masses modify the kinematics, they cannot enter in the overall prefactor $y_i^2 v^2$ that appears in the cross-sections, since the thermal masses change the dispersion relations, but preserve chirality~\cite{Bellac:2011kqa}~\footnote{One could check this by an explicit calculation using the thermal one-loop fermionic propagator, as was done {\it e.g.} in a different context in~\cite{Giudice:2003jh}, Appendix C.1. However here for simplicity we implement this by including the thermal masses in the kinematics, but not in the prefactor.}. This can also be seen by performing again the field redefinition of~\cite{Salvio:2013iaa}, which shows that the cross section must vanish for $y_i\rightarrow 0$, and therefore there cannot be a contribution to the pre-factor proportional to $m_{i, th}$, which is independent of $y_i$. The inclusion of gluon thermal masses does not change dramatically the result, but it is important conceptually, because it kinematically forbids processes with emission of many gluons, which could be otherwise dominant since we have $g_s>1$. Finally, we evaluate the couplings inside the integral at the center-of-mass energy $g_s(2 E)$ by using a simple one-loop RGE running.

We find that the rates are well approximated by the following fitting functions $\Gamma_i= e^{-\frac{m_i}{T}} m_i^2 c_i^2 g_s^2 T/f^2 (A_i+B_i \log(T/M_{Z}))$, where $M_Z$ is the Z-boson mass and $(A_t,B_t)=(1\times 10^{-2},-10^{-4})$, $(A_b,B_b)=(1.1\times 10^{-2},-10^{-4})$ and $(A_c,B_c)=(1.3 \times 10^{-2}, -2 \times 10^{-4})$ .

\section{Scattering Rates and $N_\eff$: Results}

\begin{figure}
	\includegraphics[width=7.6cm]{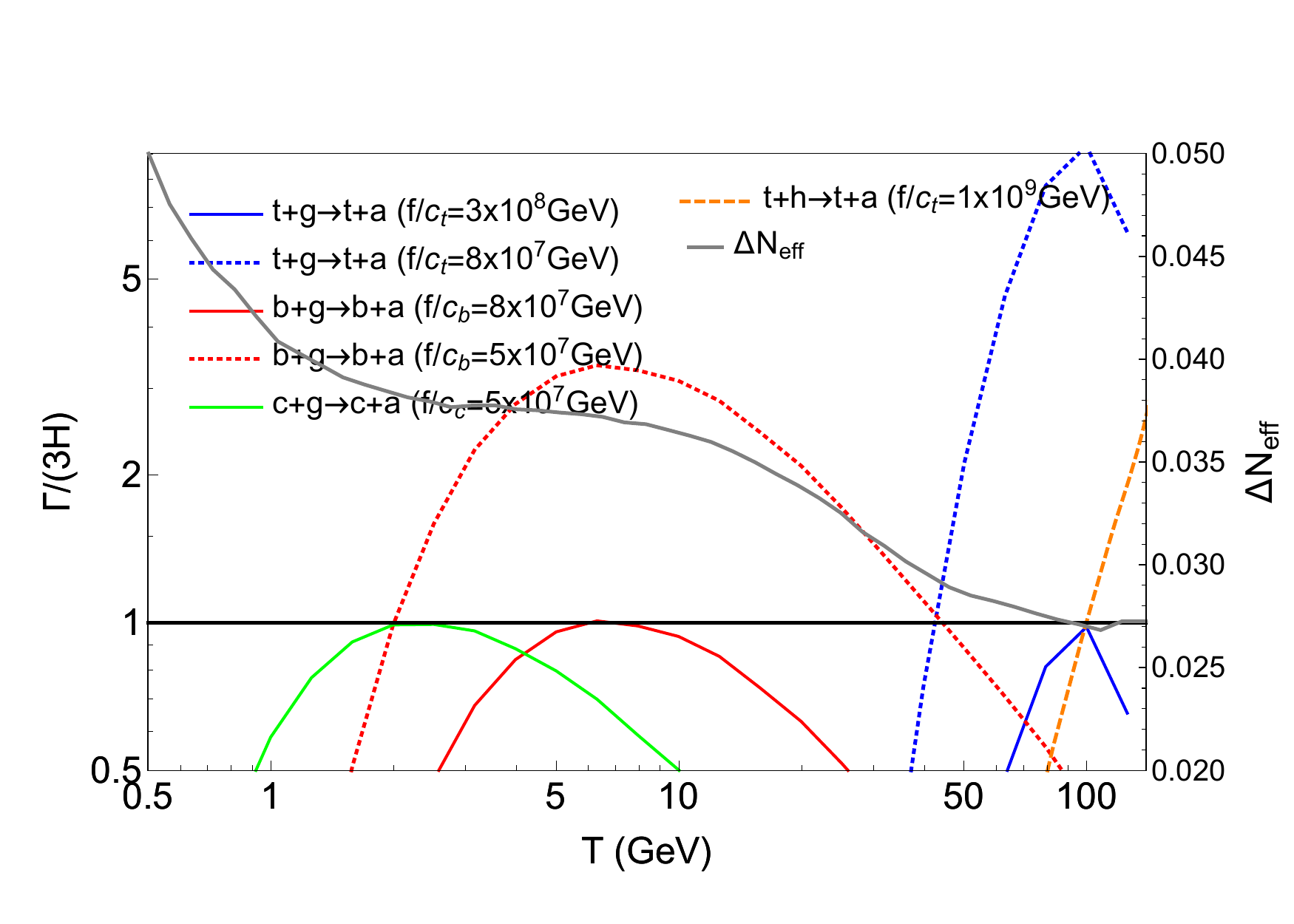}
\caption{$\Gamma_i/(3H)$ vs temperature associated with the axion production via scatterings with different quarks and gluons. Each process is evaluated at values of $f/c_i$ chosen to be the minimal ones such that $\Gamma_i/(3H)=1$. The orange line uses the result derived in \cite{Salvio:2013iaa}.  \label{Decay Rates plot}}
\end{figure}

In fig.~\ref{Decay Rates plot} we plot $\Gamma_i/(3H)$ as a function of $T$, for the scatterings involving $t$, $b$ and $c$ quarks~\footnote{We do not consider the $s$-quark, since the calculation would be unreliable, being too close to the QCDPT.}, plotting each curve at two different values of $f/c_i$. Note that what determines the final axion abundance is the decoupling temperature. In all our plots we use the one-loop analytical expression for the temperature dependence of the Higgs vev $v(T)$, which is relevant close to $100$ GeV, especially in the case of the $t$-quark. 
At large values of $f$, say $f/c_t\gtrsim10^9$ GeV, the axion can become thermal via the top-Higgs scatterings~\footnote{We use as a crude estimate the expression given in~\cite{Salvio:2013iaa}, times a Boltzmann factor $e^{-m_t/T}$, although such results were derived for massless particles, with no Higgs vev.} already at $T \gtrsim {\cal O} (100)$ GeV. However, the axion can also be produced below the EWPT through the scatterings with the $t$-quark, eq.~(\ref{Scatterings}), for $f/c_t \lesssim 10^9$GeV or, for even smaller $f$, through scatterings with the $b$ or $c$ quarks.

The important consequence of axion production is the observable effect on the effective number of relativistic degrees of freedom $N_\eff$. In the SM  this is predicted to be~\cite{Mangano:2005cc, deSalas:2016ztq} $N_\eff= 3.045$, which tells us that the only relativistic species at recombination, apart from photons, are the 3 neutrinos. Any deviation from this would tell us something new about the universe, in particular, about light relics and/or additional degrees of freedom.

A simple way to estimate $\Delta N_\text{eff} $ would be simply to read the decoupling temperature from fig.~\ref{Decay Rates plot}, and to evaluate $g_{* D}$, {\it i.e.} from~\cite{Borsanyi:2016ksw}. In order to compute $\Delta N_\text{eff}$ more precisely we solve the Boltzmann equation\footnote{This equation is valid in the Boltzmann approximation.} for the axion number density $n_a$,
\begin{eqnarray}
\frac{dn_a}{dt}+3 H n_a= \Gamma_i (n_a-n_{\rm eq}) \, .
\end{eqnarray}
 This equation can also be rewritten in terms of the abundance $Y\equiv n_a/s$, 
\begin{eqnarray}
s H z \frac{d Y}{dz}= \gamma_i \left(1-\frac{Y}{Y_{\rm eq}}\right) \, ,
\end{eqnarray} 
where $z\equiv m_i/T$, $Y_{\rm eq}\equiv n_{\rm eq}/s$, $\gamma_i\equiv \Gamma_i n_{\rm eq}$ and $s\equiv 2 \pi^2 g_* T^3/45$ is the entropy density.
\begin{figure}
	\includegraphics[width=7.2cm]{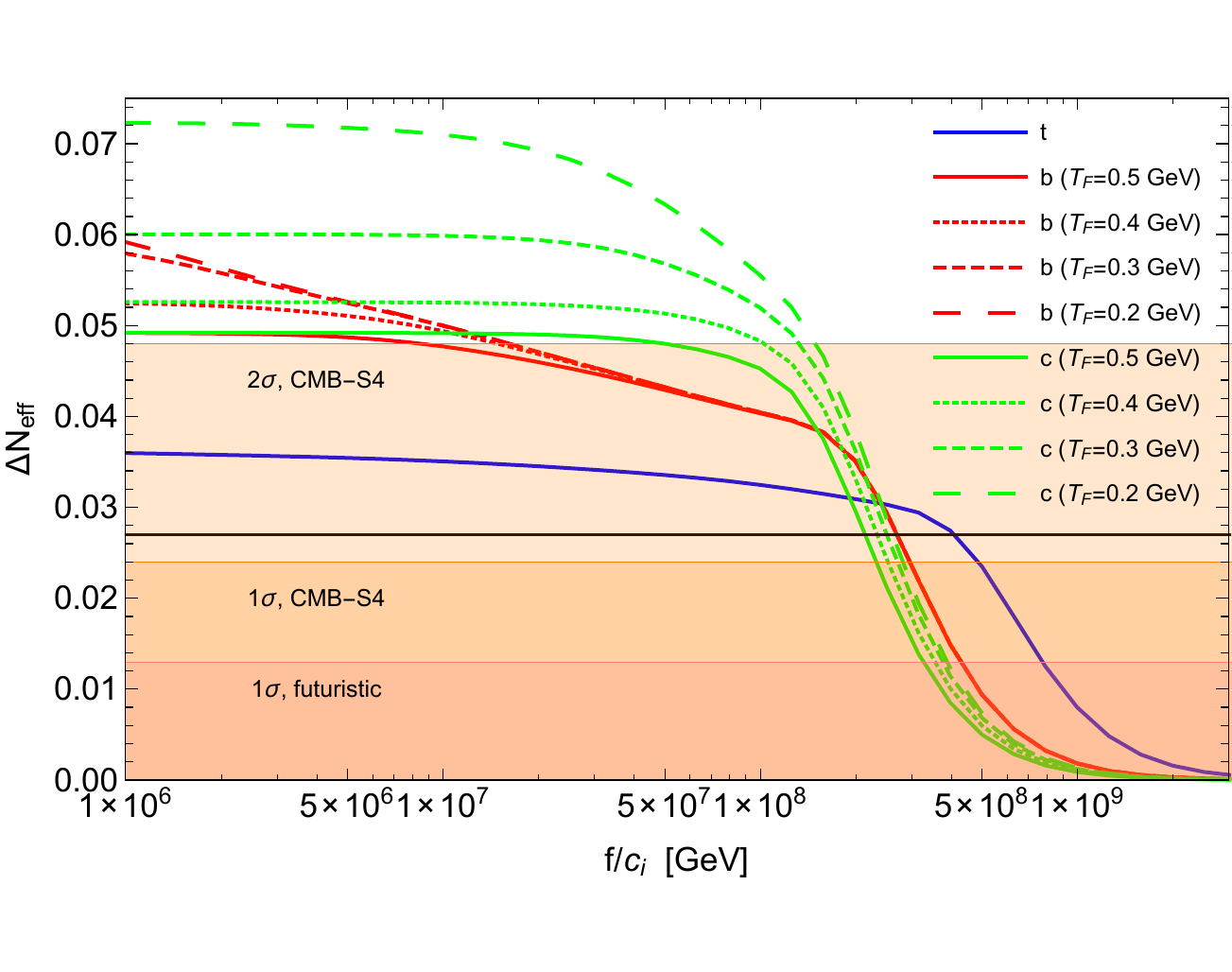}
	\caption{$\Delta N_\text{eff}$ as a function of $f/c_i$. The green, red and blue label correspond to the predictions for the charm, bottom and top particle, respectively. The orange bands represent the $1\sigma$ and $2 \sigma$ forecasted contours of CMBS4, plus a more futuristic $1\sigma$ band, according to~\cite{Abazajian:2016yjj}. See also~\cite{Baumann:2017gkg} for similar forecasts combining CMB-S4 and large scale structure.  \label{fig2}}.
\end{figure}
Assuming a constant $g_*$ there is an analytical solution. Parameterizing $\gamma_i \equiv \gamma_{m_i} z^{-4} e^{-z}$, $H \equiv  H_{m_i} z^{-2}$ and $s \equiv  s_{m_i} z^{-3}$ and imposing $Y=0$ at $z\rightarrow 0$ we get:
\begin{eqnarray}
Y=Y^{\rm eq} \left[1-e^{-(1-e^{-z}) r}\right] \, ,
\end{eqnarray} 
where $ r\equiv \gamma_{m_i}/(H_{m_i} s_{m_i} Y^{\rm eq})=(\Gamma/H)_{T=m_i}$. So, the final abundance is $Y(z=\infty) = Y^{\rm eq} (1-e^{-r})$ which corresponds to a change in $N_\text{eff}$ given by
\begin{eqnarray}
\Delta N_{\rm eff}=\frac{4}{7} \, \left(\frac{43}{4g_{*, D}}\right)^{\frac{4}{3}} \left[1-e^{-\frac{\Gamma}{H}\big \rvert_{T=m_i}}\right] \, .
\end{eqnarray}
However, $g_*$ has a non-trivial dependence with $T$~\cite{Borsanyi:2016ksw}, so we need to solve the Boltzmann equation numerically~\footnote{Other temperature dependences that we also included numerically are:  the running of $g_s$, the Higgs vev $v(T)$ and the logarithmic part of the rates $\Gamma_i$.}. 
%We use the decay rates given in the previous section and consider values of $f/c_i$ in the window $10^7<f/c_i \, \text{GeV}^{-1}<5\times10^9$. 
We stop the simulation at $T=0.5$ GeV, when $\alpha_s =1$. In the case of $t$ the results are insensitive to this cutoff. In the case of $b$ the results are also insensitive to it, provided $f/c_b\gtrsim 10^7$ GeV.  However, if the axion production is still non-zero at $T\lesssim 0.5$ GeV, as happens for the $b$, when $f/c_b\lesssim 10^7$GeV, and for the $c$, when $f/c_c\lesssim 2\times 10^8$GeV, that can affect $\Delta N_\text{eff}$, given that $g_*(T)$ varies rapidly in that region. To get a good estimate one would need more refined techniques, which we leave for future work. 

We plot the results in fig.~\ref{fig2}, showing the predicted $\Delta N_\text{eff}$ as a function of $f/c_i$ for $i=t,b,c$. As anticipated from fig.~\ref{Decay Rates plot} thermalization via top scatterings happens when $f/c_t \approx 4 \times 10^8$GeV,  leading to $\Delta N_\text{eff} \approx 0.03$.  For smaller $f/c_t$ the axion remains thermal for longer time, leading to a larger $\dn$. For example, at $10^6 {\rm GeV}\leq f/c_t\leq 10^7 {\rm GeV}$, $\dn= 0.035-0.036$. Note, however, that such values of the axial couplings to the top might already be excluded by stellar cooling constraints, $f/c_t \gtrsim 2.4\times 10^9$GeV~\cite{Feng:1997tn}. 

Similarly, the scatterings with the bottom are able to thermalize the axion at $f/c_b \approx (2-3) \times 10^8$ GeV, leading to $\dn=0.038$. In this case, at $f/c_b=10^7$ GeV, one has $\dn =0.047$ which is interestingly close to the forecasted $2\sigma$ contour of CMB-S4~\cite{Abazajian:2016yjj}. For $f/c_b$ below $10^7$ GeV we encounter the problem that we have stressed before: the axion decouples at $T\lesssim 0.5$ GeV, where the methods used here break down. To see this sensitivity we have plotted 4 red lines. The lowest one corresponds to a case where we stop the simulation at a final temperature $T_F=0.5$ GeV. Such line should be seen as a lower bound on $\dn$ at any given value of $f/c_b$. The other 3 lines correspond to cases where we  fix by hand $\alpha_s=1$ for $T<0.5$ GeV and then run the simulation down to lower temperatures $T_F={0.4, 0.3, 0.2}$ GeV, respectively.   The lowest red line  gives a lower bound on $\dn$, which at at $10^6 {\rm GeV}\leq f/c_b\leq 10^7$ GeV, is $\dn>0.049$.
 
Regarding the scatterings with the charm they are able to thermalize around $f/c_c \lesssim  10^8$ GeV. This case is affected, to a greater extent, by the same uncertainty due to strong coupling and the final predictions turn out to be very sensitive to what happens below $0.5 {\rm GeV}$. Also in this case we plot 4 lines to show the sensitivity to $T_F$ and again, from the lowest green line we can give a lower bound on $\dn$ which at $f/c_c<10^{8}$GeV is $\dn>0.045$.  Such values, obtained close to the strong coupling region, are of great interest, since they could be potentially measurable at  2$\sigma$  by CMB-S4. However, a dedicated study of such cases would be needed, using more adequate methods to treat the axion decoupling around the QCDPT. 

Finally, note that in all cases even if the axion does not thermalize completely it can still have a non-negligible abundance, and even give $\dn>0.027$.

In conclusion, we have shown that, in addition to the possible thermalization at $f \gtrsim 10^{9}$ GeV, before the EWPT via gluons~\cite{Masso:2002np,Graf:2010tv} or axion-top-Higgs scatterings~\cite{Salvio:2013iaa}, the axion can be produced at intermediate energy scales between the EWPT and the QCDPT via direct couplings to heavy quarks. Such couplings are present in various models, such as DFSZ, at tree level and thus are enhanced compared to interactions with gluons, which are generated at one loop. Clearly, because we are probing lower energies, this happens in a lower range of $f$, which by solving the associated Boltzmann equation turns out to be $ f/c_i \lesssim 10^9 \text{GeV}$. This window can be viable if SN bounds are not so robust~\cite{Giannotti:2017hny} or, as pointed out in \cite{DiLuzio:2017ogq}, in DFSZ-like axion models which can avoid SN bounds, because of suppressed coupling to light quarks and thus to nucleons. Interestingly in the same window one could also explain the hints of extra stellar energy losses, as suggested in~\cite{Giannotti:2017hny}~\footnote{Note also that, although this region does not overlap with the standard window of axion dark matter, it may be possible to obtain QCD axion DM even at such small $f$, via domain wall decay~\cite{Giannotti:2017hny,Kawasaki:2014sqa,Ringwald:2015dsf} or via resonant production of axions at the PQ phase transition~\cite{Co:2017mop}, although in such cases one should check whether the scatterings we consider can destroy the condensate. Furthermore our window of $f$ can also be relevant in the context of Axion-like particles (ALPs), as in~\cite{Daido:2017wwb}. }.

Axion production and decoupling happen at different thresholds $T_{i}$, close to the heavy quark masses $m_i$, leading to different predictions on $\Delta N_{\rm eff}$ as a function of $f/c_i$, shown in fig.~\ref{fig2}, which contains our main results. We stress that such predictions are not upper bounds, but can be precisely given as far as we know the number of degrees of freedom $g_{*, D}$ and the dynamics at such energies. For the top and bottom our main predictions ($0.027\leq \dn \leq 0.036$ for $10^6 {\rm GeV} \lesssim f/c_t\lesssim 4\times 10^8$ GeV and  $0.027\leq \dn \leq 0.047$ for $10^7 {\rm GeV}  \lesssim f/c_b\lesssim 3\times  10^8$GeV) are quite under control, while 
%For $10^7 {\rm GeV} \lesssim f/c_t\lesssim 4\times 10^8$ GeV top scatterings lead to $0.027\leq \dn \leq 0.035$, while for $10^7 {\rm GeV}  \lesssim f/c_b\lesssim 2.7\times  10^8$ GeV bottom scatterings lead to $0.027\leq \dn \leq 0.035$, which are both larger than the abundance one could get via thermalization above the EWPT. 
for the $c$-quark, since axion production is too close to the QCD phase transition, we  only have a lower bound $\dn \geq 0.045$ (at $f/c_c\lesssim 10^8$ GeV), which nonetheless might represent the largest signal for a detection.

These observations open a new window to test the QCD axion. At the same time, if the axion exists in this range of $f$, this would allow us to probe the Early universe at energies so far unexplored in cosmology, between the QCD and the EW phase transitions. If the axion could be detected directly within such window for $f$, {\it e.g.} with the CAST~\cite{Arik:2008mq} or IAXO~\cite{Armengaud:2014gea, Irastorza:2013dav, Irastorza:2011gs} experiments, our conclusions would make possible to confirm if our expectations about the total particle content and the thermal history at such temperatures are indeed correct.

	${}$\linebreak
	\emph{\textbf{Acknowledgments:}}
	We thank Luca di Luzio for initial discussions about the project, Federico Mescia, Javier Redondo, Alberto Salvio, Francesco d'Eramo, Fernando Arias-Aragon and Andreas Ringwald for enlightning discussions and useful comments. This work is supported by the grants EC FPA2010-20807-C02-02, AGAUR 2009-SGR-168, ERC Starting Grant HoloLHC-306605 and by the Spanish MINECO under MDM-2014-0369 of ICCUB (Unidad de Excelencia ``Maria de Maeztu'').

	\bibliographystyle{JHEP}
	\bibliography{QCDAxionTh.bib}
	
	%%%%
\end{document}